\documentclass[pra,aps,twocolumn,superscriptaddress,amsmath,amssymb,floatfix,showpacs]{revtex4}
\usepackage{color,graphicx,bm,amsmath}

\begin{document}

\newcommand{\half}{\ensuremath{\frac{1}{2}}}
\newcommand{\degree}{\ensuremath{^{\circ}}}
\newcommand{\xnl}{\ensuremath{\chi^{(3)}}}
\newcommand{\xnlordinary}{\ensuremath{\chi^{(2)}}}
\newcommand{\mx}{\ensuremath{\mbox{\tiny{\rm X}}}}
\newcommand{\mc}{\ensuremath{\mbox{\tiny{\rm C}}}}

\title{Effect of Interactions on Vortices in a Non-equilibrium Polariton Condensate}


\author{D.~N.~Krizhanovskii}

\affiliation{Department of Physics and Astronomy, University of
Sheffield, Sheffield, S3 7RH, UK}

\author{D. M. Whittaker}

\affiliation{Department of Physics and Astronomy, University of
Sheffield, Sheffield, S3 7RH, UK}

\author{R. A. Bradley}

\affiliation{Department of Physics and Astronomy, University of
Sheffield, Sheffield, S3 7RH, UK}

\author{K. Guda}

\affiliation{Department of Physics and Astronomy, University of
Sheffield, Sheffield, S3 7RH, UK}

\author{D. Sarkar}

\affiliation{Department of Physics and Astronomy, University of
Sheffield, Sheffield, S3 7RH, UK}

\author{D. Sanvitto}

\affiliation{Departmento Fisica de Materiales, Universidad Autonoma de Madrid, 28049 Madrid, Spain}

\author{L.Vina}

\affiliation{Departmento Fisica de Materiales, Universidad Autonoma de Madrid, 28049 Madrid, Spain}

\author{E. Cerda}

\affiliation{Paul-Drude-Institut f\"ur Festk\"orperelektronik, Berlin, Germany}

\author{P. Santos}

\affiliation{Paul-Drude-Institut f\"ur Festk\"orperelektronik, Berlin, Germany}

\author{K. Biermann}

\affiliation{Paul-Drude-Institut f\"ur Festk\"orperelektronik, Berlin, Germany}

\author{R. Hey}

\affiliation{Paul-Drude-Institut f\"ur Festk\"orperelektronik, Berlin, Germany}

\author{M. S. Skolnick}

\affiliation{Department of Physics and Astronomy, University of
Sheffield, Sheffield, S3 7RH, UK}

\date{\today}

\begin{abstract} 

We demonstrate the creation of vortices in a
macroscopically occupied polariton state formed in a semiconductor
microcavity. A weak external laser beam carrying orbital angular
momentum (OAM) is used to imprint a vortex on the condensate arising from the polariton
optical parametric oscillator (OPO). The vortex core
radius is found to decrease with increasing pump power, and is
determined by polariton-polariton interactions. As a result of OAM conservation in the parametric scattering process, the excitation consists of a vortex in the signal and a corresponding
anti-vortex in the idler of the OPO.  The experimental results are in good
agreement with a theoretical model of a vortex in the polariton OPO.

\end{abstract}


\maketitle

Microcavity polaritons are bosonic quasiparticles, which arise from
strong exciton-photon coupling in semiconductor microcavities. Characteristic bosonic phenomena, such as stimulated scattering ~\cite{stevenson2000} and  polariton condensation have been reported~\cite{Kasprzak2006}. In contrast to well-known
atom Bose-Einstein condensates (BEC), polariton condensates are intrinsically out-of-equilibrium:
a balance is reached between external excitation and losses due to the
finite polariton lifetime\cite{WoutersPRL,KrizhPRB2009}. In this respect the system has similarities to a laser, but with the fundamental difference that the particles have hybrid light-matter character with strong interactions due to the excitonic component.

Non-equilibrium
dynamics can lead to a much richer phenomenology of spontaneous
pattern formation than in equilibrium systems\cite{cross_hohenberg}, with vortices a typical example. 
Vortices are found in many areas of science, including particle
physics, optics and condensed matter systems.  They are characterised
by winding of phase around a point, the vortex core, where the density
of the system goes to zero. The total phase change for a complete loop
must be an integer multiple of $2\pi$, so a vortex is a state
with quantised orbital angular momentum (OAM). 
Vortices have been intensively studied in classical systems including optical vortex beams ~\cite{vortex-beam}, and lasers ~\cite{laser-science}. They are also a
characteristic property of condensed-phase systems like BEC, liquid helium and superconductors ~\cite{pitaevski}.

Spontaneous formation of vortices has been observed in an incoherently
pumped polariton system~\cite{lagoudakis}.  In this paper we
demonstrate that vortices in a polariton condensate can be created 
using a weak external imprinting beam (which we term the \textit{im}-beam).  We find that the vortex core radius is determined by polariton-polariton interactions and
show that these lead to a decrease of the radius with increasing particle
density. Here, the low polariton mass ($\sim 10^{-5}m_e$) plays an important role;
the polariton vortex radius is typically $\sim 10
\mu\mbox{m}$, some 2 orders of magnitude larger than in an atom
system. This enables us to observe in situ the vortex profile
and its density dependence, unlike in atom BEC experiments where vortices can only
be imaged following condensate expansion\cite{madison}, which is a destructive technique. 


The system we study is a microcavity optical parametric oscillator
(OPO)~\cite{stevenson2000}, in which coherent high density signal (s)
and idler (i) states (condensates) arise from polariton-polariton scattering
from the pump (p), as shown in Fig. 1(a). An important property of the
OPO is that the phase of the signal and idler are not determined by
that of the pump~\cite{Wouters2007}, but appear as a spontaneous symmetry breaking
at the OPO threshold, similar to the phase transition
of a thermal equilibrium BEC. Both the OPO modes, and the polariton
condensate which is realised by non-resonant pumping
\cite{Kasprzak2006} exhibit characteristic properties of a
system possessing a macroscopic wavefunction, such as extended spatial
and temporal coherence~\cite{Kasprzak2006, Richard2006, Love2008,Krizh2006}.  

As well as vortex creation we demonstrate OAM conservation in the parametric scattering process; this means that
when a vortex is imprinted onto the OPO signal, polariton-polariton scattering results in an anti-vortex with opposite
OAM in the idler, as has been predicted previously theoretically \cite{whittaker2007}. It is this vortex/anti-vortex pair that is created
in our experiment. We show theoretically that when the signal population is
small compared to the pump, which is the regime of the present
experiments, the signal and idler vortices have identical
profiles, determined by the interactions.


The experiments were carried out at 4K on a 
GaAs-based microcavity grown by molecular beam epitaxy. It has a Rabi
splitting of $6$ meV and zero detuning between the quantum well
exciton and the optical mode.  A single mode Ti-Al$_2$O$_3$ laser on resonance
with the lower polariton branch at $14^0$ was used to excite the
OPO (Fig. 1(a)). The zero OAM pump was focused to a $\sim50$ $\mu$m
Gaussian spot.  A spatial light modulator was used to generate a
Gauss-Laguerre optical beam, corresponding to an optical vortex with
OAM $m=1$ which was used as the imprinting beam.

\begin{figure}
\begin{center}
\mbox{
\includegraphics[scale=0.55]{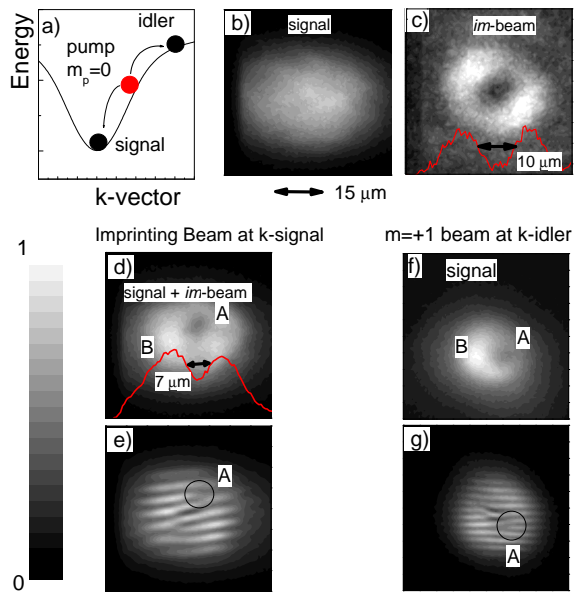}
}
\end{center}
\caption{
(a) Schematic diagram of OPO. (b) Real space image of the signal with
no imprinting beam. (c) Image of Gauss-Laguerre imprinting beam (\textit{im}-beam). (d) Signal with weak \textit{im}-beam of (c), showing an imprinted vortex labeled A. (e)
Interferogram revealing the $2\pi$ phase variation
around the vortex of (d). 
(f) Vortex in signal, labeled A, created by excitation at
the idler position with Gauss-Laguerre beam. (g) Interferogram revealing
the vortex with OAM $m_s =-1$ created in (f). The
red lines are cross-sections through the vortex cores of (c) to (d),
with the sizes of the cores (FWHM) indicated.}
\end{figure}

Fig. 1(b) shows the spatial image of the OPO signal emission, which
occurs at $1.533$ eV and has a line-width of $<60$ $\mu$eV (resolution
limited), at a pump power of $~3$ times the OPO threshold, $P_{\rm
th}\sim600$ $\mbox{Wcm}^{-2}$. It has a uniform spatial distribution,
consistent with a sample having very weak disorder
~\cite{sanvitto2006}. In order to create a vortex in the signal we
introduce a weak \textit{im}-beam of power $\approx 10$
$\mu$W, $\sim40$ times less than that of the signal, having OAM $M=+1$ and optical vortex core diameter of $10$ $\mu$m. The
\textit{im}-beam is in resonance with the signal at normal
incidence. Application of the \textit{im}-beam of Fig. 1(c) results
in a strong modification of the spatial distribution of the signal
(Fig. 1(d)), characterized by a resultant well-defined dip of diameter
(FWHM) $\approx 7 \mu$m, (radius $\approx 3.5 \mu$m (HWHM)) labeled A on Fig. 1(d).

To demonstrate that the dip arises from a vortex, the spatial phase
variation of the signal was measured \cite{lagoudakis}. 
Fig. 1(e) shows the interference pattern between the signal image in
Fig. 1(d) and the image inverted around a central point of symmetry so
that region $A$ interferes with region $B$ (see Fig. 1(d)), where the
phase is nearly constant. The two fork-like dislocations in the
interference pattern demonstrate the presence of a single vortex, with
phase winding by $2\pi$ about the core. One fork occurs at A, the
position of the vortex in the original beam, while the other occurs at
the equivalent location in the inverted beam.

We describe the process using the term `imprinting' because the
\textit{im}-beam is very weak; we are imaging not the polaritons injected
directly by the \textit{im}-beam , but the OPO signal modified by their
presence. Indeed, the vortex disappears when the \textit{im}-beam is tuned off
resonance 
with the signal. The reason imprinting can occur is that the
phase of the signal is undetermined in the OPO, so even the few
polaritons the \textit{im}-beam injects are able to lock the signal phase to
their own. Experimentally there is a minimum ratio of
the \textit{im}-beam to signal power density, of about $\sim 1/45$, required to create a
vortex in the signal phase. This limit is probably determined by the
fluctuations which lead to decoherence;\cite{Love2008,Krizh2006} at
lower powers the \textit{im}-beam is unable to overcome their effects.

The fact that the dip in the vortex core does not go fully to zero
(the intensity typically reduces by a factor of $\sim 1.6$) suggests
that the observed signal is not a simple OAM state with $m_s=1$, but
contains also some background with $m_s=0$.  This probably arises from
unresolved multi-mode spectral structure in the signal, with the
vortex only imprinted on the stronger features.

\begin{figure}
\begin{center}
\mbox{
\includegraphics[scale=0.35]{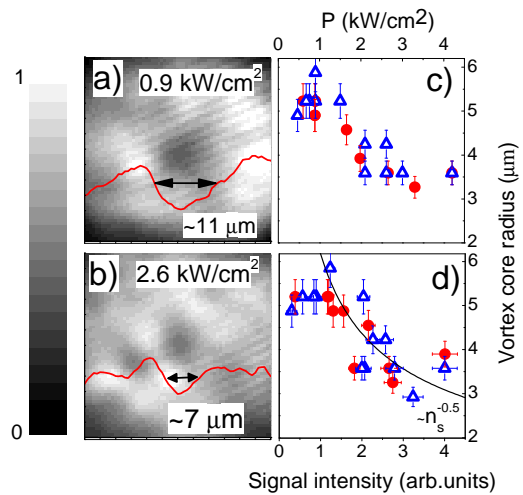}
}
\end{center}
\caption{
a)-b) Images of a signal vortex at (a) $0.9$ $\mbox{kWcm}^{-2}$ 
and (b) $2.6$ $\mbox{kWcm}^{-2}$. Experimental vortex core radius
of the OPO signal as a function of pump power (c) and as a function of the signal intensity (d) for an \textit{im}-beam  core radius of $\sim 4\mu$m (circles) and $\sim 7\mu$m (triangles), respectively.}
\end{figure}

Vortices and OAM conservation have previously been investigated in
nonlinear optics using down-conversion
experiments\cite{Martinelli}. However, the non-linearity in
those works has a $\chi^{(2)}$ form. A fundamental difference in the
polariton system is that the non-linearity is $\chi^{(3)}$. This means
that, as well as giving rise to the OPO process, it leads to self (and
mutual) interaction terms, which give rise to the blue-shifts of the modes \cite{whittaker2001}. 
These interaction terms are not found in $\chi^{(2)}$ processes.
The polariton vortex profiles are expected to be determined by
these interactions, with a characteristic length scale, the healing length,
which depends on the population, as we show below.

In order to investigate this population dependence we have measured the vortex radius at different pump
powers. Since the signal population increases with the pump power the \textit{im}-beam  power was adjusted to keep the ratio of the signal to \textit{im}-beam  power density constant at $\sim15$.  
Figs. 2(a) and (b) show the vortex image at P=0.9
$\mbox{kWcm}^{-2}$ $\sim 1.5 P_{th}$ and P=2.6 $\mbox{kWcm}^{-2}$ $\sim
4.5P_{th}$, respectively. The reduction of the vortex size with
increasing excitation power is apparent. Figure 2 (c) shows the
dependence of the vortex core radius on excitation power; a
decrease from $\sim5.5$ $\mu$m at threshold down to $\sim3$ $\mu$m at
excitation density 5 times above threshold is observed. Moreover, very similar
vortex sizes and variation with polariton density were obtained for two different sizes of the \textit{im}-beam  with vortex core radius of $\sim 4\mu$m (circles) and $\sim 7\mu$m (triangles). These results
show that the profile of the imprinted vortex is an intrinsic property of the interacting polariton system.

The theory below shows that the vortex radius is given by the healing length $\xi=(2 M_c \kappa
\sqrt{n_s n_i})^{-\frac{1}{2}}$, where $M_c$ is the cavity photon effective mass,
$\kappa$ the strength of the non-linearity, and $n_s$, $n_i$ the
signal and idler population densities. 
This can be understood qualitatively using the following argument, which applies to  interacting equilibrium BECs \cite{pitaevski}, and to lasers (VCSELs) with  $\chi^{(3)}$ nonlinearity \cite{laser-science};
the healing length is obtained by equating a typical kinetic energy associated with a vortex in the condensate, $\sim (2 M_c \xi^2)^{-1}$ to the interaction energy (blueshift), which is  $\approx \kappa \sqrt{n_s n_i}$ (all energies are scaled to $\hbar^2$).
 Since $n_s$ is proportional to $n_i$ we expect the vortex radius to be approximately proportional to $n_s^{-\frac{1}{2}}$. 
 Fig. 2(d) shows that at higher signal intensities the experimental data are in very reasonable agreement with this dependence (solid line); with increasing signal intensity by a factor of 4 the vortex radius decreases by a factor of 2. At small signal intensities the vortex radius does not diverge as expected from the $n_s^{-\frac{1}{2}}$, but is probably limited by the finite signal size.

\begin{figure}
\begin{center}
\mbox{
\includegraphics[scale=0.38]{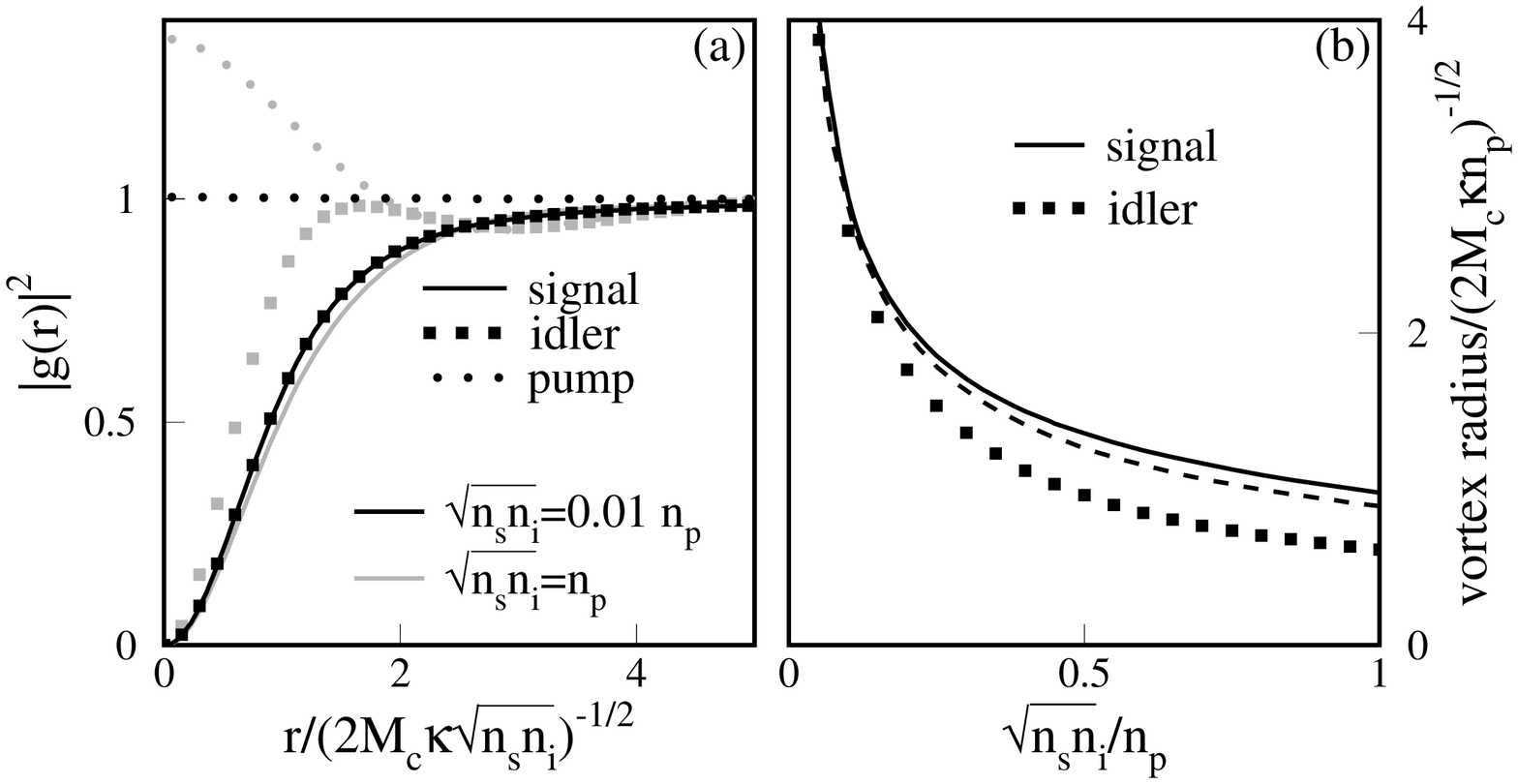 }
}
\end{center}
\caption{
(a) Calculated signal (solid line), idler (square) and pump (dot) vortex profiles for signal and idler
populations near threshold (black), and comparable to the pump (grey).  
(b) Variation of signal (solid line) and idler (square) core radii with signal/idler population.
The dashed line is obtained from the near-threshold  approximation Eq.(1).}
\end{figure}

Experimental constraints hinder the direct
demonstration of the creation of the anti-vortex in the idler accompanying
the signal vortex; the OPO idler is typically $>100$ times weaker than
the signal \cite{Krizh2006} due to the small photonic component at the idler, making it impossible to obtain spatial and
interference images.  However, we are able to show that the pair
scattering leads to the creation of vortex anti-vortex
pairs. Using a pump-power just below the OPO threshold, we apply a
beam with OAM $M=+1$ at the angle and energy where the OPO idler would
appear. The `signal' then forms at $k \sim 0$ due to pair
scattering from the pump, stimulated by the 
probe beam. In this geometry, we are able to perform the
required imaging of the signal; the vortex image and the corresponding self-
interference image are shown in Figs. 1(f) and (g),
respectively. A dip in intensity  is observed in
Fig. 1(f) corresponding to a vortex core. In Fig. 1(g) a fork
dislocation corresponding to a vortex labeled A  in
Fig. 1(f)  is observed, pointing to the right rather than
to the left as for the vortex labelled A in Figs. 1(d,e), proving that the OAM is $m_s=-1$. Here, in Fig. 1(f) we image the same original (uninverted) beam as in Fig.1 (d). 

In order to obtain a better understanding of the nature of the vortex,
we have developed a theoretical model of a vortex in a $\chi^{(3)}$\/
OPO. Our treatment starts from the classical model of an OPO with
uniform fields for pump, signal and idler,\cite{whittaker2001} and adds
a vortex anti-vortex pair to the signal and idler, which locally
modulates the fields in the region of the core. The modulation
functions are determined by a set of coupled non-linear differential
equations, which we solve numerically. Details of the theory will be
presented elsewhere. 

Fig. 3(a) shows typical profiles calculated for the three modes at
low ($\sqrt{n_s n_i}=0.01 n_p$) and high ($\sqrt{n_s n_i}=n_p$) signal
intensities. Here $n_p$ is the polariton density in the pump, which is nearly constant above threshold \cite{whittaker2001}.
 The calculation uses structural parameters appropriate to the
experimental microcavity. The length scale in the plot is the healing length $\xi=
(2M_c\kappa \sqrt{n_s n_i})^{-\frac{1}{2}}$. At small pump powers when $\sqrt{n_s n_i}=0.01 n_p$ the signal and idler profiles are identical with radii of $\sim \xi$, and the pump is unaffected by the vortex.   
At high powers,
the signal and idler profiles separate, with the idler having the
smaller core, and a `bump' appears in the pump in the core region, where
it is less depleted by the OPO process. The separation 
 is shown in more detail in Fig. 3(b), where
the vortex radii are plotted as a
function of $\sqrt{n_s n_i}/n_p$. In this figure, the length scaling
used is $(2M_c\kappa n_p)^{-\frac{1}{2}}$, to show explicitly the strong 
dependence on population.

The behaviour can be understood using physical arguments
based on the competition between the blue-shift due to polariton-polariton interactions and the OPO scattering process
arising from the non-linearity\cite{whittaker2001}. The blue-shift provides restoring
forces which control the vortex radius, as in the standard
Gross-Pitaevskii equation\cite{pitaevski}. It tends, at high powers, to produce
different radii for the signal and idler because of the different
masses; the more excitonic idler has a larger mass and so
a smaller core. The OPO effect, by contrast, tries to make the
profiles the same, because the OPO scattering is a local process,
generating equal numbers of signal and idler polaritons at a given
point in space.

 The regime when the signal (and idler) population is much smaller than that in the pump ($n_{s}\sim 10^9-10^{10}$ cm$^{-2}$ $\leq 0.2n_p$), is the one which is accessed in the present experiment \cite {Krizh2006}. In this case the
OPO process dominates and the signal and idler profiles are locked
together. The actual profile, $g(r)$, is, however, still determined by
the blue-shift terms; we can show that it satisfies an equation of
the complex Ginzburg-Landau (CGL) type. Scaling the radius by the healing length $\xi$,
we find
\begin{align}
\left( \nabla_r^2-\frac{m^2}{r^2} \right) \, g
&=\alpha(|g|^2-1)g
\end{align}
where $\alpha$ is a complex constant with magnitude of order unity.
It is determined by the ratio of the signal and idler line-widths, the
exciton and cavity amplitudes (Hopfield factors) for the modes, and
the phase difference $\Phi=2
\phi_p-\phi_s-\phi_i$ between the pump, signal
and idler phases, respectively.
The solutions of this equation provide the dashed curve on Fig. 3(b),
which is in very good agreement with the full theory for $\sqrt{n_s n_i}
\lesssim 0.1 n_p$.
We note that a CGL equation of the same form is obtained for the
incoherently pumped polariton condensate.\cite{KeelingPRL}
Although the OPO  is a more complicated system, with three
coherent fields, in the near threshold regime it displays similar
physics.


In weak coupling microcavity lasers (VCSELs), which operate at one to two order of magnitude higher electron-hole density than the polariton condensate, there is also an effective $\chi^{(3)}$ nonlinearity due to saturation of the gain \cite{laser-science}. This contrasts with the direct interactions between polaritons due to their exciton component. At low powers the VCSEL can be approximated by a CGL type equation similar to Eq.1, and hence a decrease of vortex size with power is expected. However, the variation is expected to be more complicated as saturation is approached. A typical vortex size of several microns is observed in Ref.\cite{laser-science}, but the physical origin of this is not discussed. To the best of our knowledge there have not been any studies on the variation of vortex size with pump power in VCSELs. Instead as the current density was increased in Ref. \cite{laser-science}, spontaneous vortex proliferation was observed.

It is also interesting to compare the magnitudes of the vortex radius in the 
polariton and atom systems: for
polaritons, $M_c \sim 10^{-5}m_e$, and $\kappa \sqrt{n_s n_i} \sim
10^{-1}$ $\mbox{meV}$. 
Equivalent values for a rubidium BEC with a density
$n\sim10^{14}$ $\mbox{cm}^{-3}$ would be $M \sim 10^{5} m_e$ and $g n=4 \pi
a_s n \hbar^2/M \sim 10^{-7}\mbox{meV}$, where the atom scattering length $a_s\approx5$ nm \cite{pitaevski} and g is the atom interaction strength. This gives radii
for the polariton and the atom systems of order 10 $\mu$m and 0.1 $\mu$m
respectively.   We see that the much smaller polariton mass compensates the stronger interactions, giving a healing length two orders of magnitude greater than atom BECs.


To conclude, the polariton OPO supports a novel
excitation consisting of a vortex in the signal and an
anti-vortex in the idler. We have shown that the core radius of vortices in the polariton condensate is determined by polariton-polariton interactions.  With ultra-fast
measurement techniques the imprinting method also provides a means to investigate vortex
dynamics on time-scales inaccessible in other systems.  By including
polarisation as well as phase variation in the imprinting beam, it
should be possible to study more complicated topological defects, such as `half-vortices' ~\cite{Rubo2007}.

This work was supported by grants EP/G001642,
EP/E051448, MEC MAT2008-01555/NAN and QOIT-CSD2006-00019. We thank M. Wouters, Z. Hadzibabic and K. Burnett for helpful discussions.

\end{document}